\documentclass[twocolumn,english,showpacs,preprintnumbers,amsmath,amssymb]{revtex4}
\usepackage{float}
\usepackage{graphicx}
\usepackage{dcolumn}
\usepackage{bm}
\makeatletter
\usepackage{babel}
\makeatother

\begin{document}

\title{Stability of Bose-Einstein condensates in a circular array}
\author{E. T. D. Matsushita and E. J. V. de Passos}
\affiliation{Instituto de F\'{i}sica, Universidade de S\~{a}o Paulo, Caixa Postal 66318,
C.E.P. 05389-970, S\~{a}o Paulo, S\~{a}o Paulo, Brazil}

\begin{abstract}
The properties of the superfluid phase of ultra cold bosonic atoms
loaded in a circular array are investigated in the framework of the Bose-Hubbard model and the Bogoliubov theory. We derive and solve the Gross-Pitaevskii equation of the model to find that the atoms condense in states of well-defined
quasimomentum. A detailed analysis of the coupling structure in the effective quadratic grand-canonical Hamiltonian shows that only pairs of distinct and identical quasimomenta are coupled. Solving the corresponding Bogoliubov-de Gennes equations we see that each pair of distinct quasimomenta gives raise to doublets in the excitation energy spectrum and that the quasimomenta of the zero-energy mode and of the occupied state in the condensates are identical. The dynamical and energetic stabilities of the condensates are determined by studying the behavior of the elementary excitations in the control parameters space. Our investigation establishes that superflow condensates exists only in the central region of the first Brillouin zone whereas there is none in the last quarters since they are energetically unstable, independently of the control parameters.
\end{abstract}
\maketitle

\section{Introduction}

The experimental realization of Bose-Einstein condensates in a periodic
potential created by interference of laser beams, called an optical
lattice \cite{Raithel,Muller,Friebel}, opened up the possibility
of studying the properties of ultra cold atoms in a variety of conditions.
Examples are the access to strongly interacting atomic regimes \cite{Chu,Kasevich}
and to states which are fundamental for applications in quantum information
\cite{Jaksch}. These systems have proven to be a rich research field
that offer the unique possibility to investigate fundamental questions
from condensed matter physics to quantum optics. 

A Bose-Einstein condensate in an optical lattice is nearly a perfect
physical realization of the Bose-Hubbard model, a fact first pointed
out by Jaksch and coworkers \cite{Jaksch_Bruder}. This model gives
a description of the physics of interacting bosonic atoms trapped
in a lattice potential with atomic tunneling between nearest neighbors
and on-site repulsion. The main advantage of these systems is that
they are highly controllable. As an example, when the ratio between
the interaction and hopping terms is varied by changing the laser
intensity, a quantum phase transition from a superfluid phase to a
Mott insulator phase is observed \cite{Greiner}.

In this paper we use the Bose-Hubbard model and the Bogoliubov theory to study the dynamical
and energetic stabilities of Bose-Einstein condensates loaded in a
periodic ring. The dynamical and energetic stability of condensates in optical lattices have been investigated both theoretically \cite{Stringari,Paraoanu,Wu,Machholm,Modugno} and experimentally \cite{Fallani,DeSarlo,Campbell,Mun,Bloch}. The purpose of this paper is to give a complete and self-contained
description of the properties of condensates, its excitation spectrum
and stability. Briefly, we characterize the condensates
by the quasimomenta of occupied states which reveals the existence
of equilibrium current carrying condensates \cite{Paraoanu,Bloch}, we analyze the coupling structure
of the effective quadratic grand-canonical Hamiltonian in the quasimomenta
basis which allowed us to establish the composition of the elementary
excitations, the doublet structure of excitation energy branch and
the proper assignment of the zero-energy mode \cite{Blaizot} and the phonon limit. We find that in the zero-energy mode, which is a direct consequence of the atom number violation by the Bogoliubov theory, the quasimomenta of the atoms are identical to the quasimomentum of the occupied state in the condensate. Turning to the phonon limit we find that it is achieved when the relative quasimomentum $l-p$ goes to zero, with $l$ and $p$ being, respectively, the quasimomenta of the excitation and of the occupied state in the condensate, a fact overlooked in the literature \cite{Paraoanu,Smerzi}.

The stability of the condensates is determined by the behavior
of the excitation energies and composition of the elementary excitations
in the control parameters space. In fact an equilibrium state is dynamically
stable if all excitation energies are real, \cite{Paraoanu,Machholm,Wu,Modugno}, and it is energetically stable if they are positive, \cite{Paraoanu,Fetter,Machholm,Modugno}. In our study we identify two mechanisms
of energetic instability: ``crossing" that occurs when the excitation
energy vanishes and change its sign and ``no-crossing" that occurs
when the excitation energy is strictly negative, independently of the control parameters. From this analysis we determine the domains in the control parameters space where it is
possible to find metastability in the system, that is, dynamically and
energetically stable condensates. Metastable current carrying condensates correspond to local minima of the energy and they are candidates to present superfluid motion \cite{Wu,Bloch}.

This paper is organized as follows. In the Section II we derive and
solve the Gross-Pitaevskii equation for the Bose-Hubbard model to
show that the atoms condense in states with well-defined quasimomentum \cite{Paraoanu}, revealing the existence of equilibrium current carrying states \cite{Bloch}. Besides, these
states with well-defined quasimomentum define a single-particle basis
which diagonalizes the hopping term of the Bose-Hubbard Hamiltonian.
In the Section III we determine the energies and the composition of
the elementary excitations by the diagonalization of the effective
grand-canonical Hamiltonian of the Bogoliubov theory. We express it in the quasimomentum representation
to show that only pairs of quasimomenta are coupled. We identify these
pairs to cast the effective grand-canonical Hamiltonian into the form
of a sum of terms each one involving pairs of identical and distinct
quasimomenta. This fact allow us to reduce the process of diagonalization
of a $2M\times2M$ matrix, with $M$ being the number of lattice sites,
to one of $2\times2$ matrices when the quasimomenta of the pairs
are identical and $4\times4$ matrices when they are distinct, which
implies that the effective grand-canonical Hamiltonian is diagonal
in blocks. In fact, when $M$ is odd, we have $\frac{M-1}{2}$ pairs
with distinct quasimomenta and only one pair with identical.
On the other hand, for $M$ even, we have $\frac{M}{2}-1$ pairs with
distinct quasimomenta and two pairs with identical. We
will see that $2\times2$ blocks correspond to one excitation energy
whereas $4\times4$ blocks correspond to doublets. In the Section
IV we determine the dynamical and energetic stability of the condensates
by studying the behavior of excitation spectrum and the composition of the elementary excitations in the control parameters
space. We found that the stability properties of the condensates depend
only on a combination of control parameters $r\equiv\frac{nU}{2J}$,
where $n=\frac{N}{M}$, $J$ and $U$ are,
respectively, the density of atoms, the hopping and the on-site strenghts. We determine
critical values of $r$ that define, in the control parameters space,
the domains where it is possible to find superflow states in the
system. The dynamical and energetic phase diagrams are shown. A summary and our conclusions are presented in the Section V.

\section{The Gross-Pitaevskii equation of the Bose-Hubbard model}

\subsection{The Bose-Hubbard Hamiltonian}

The physics of an ultra cold, dilute and interacting Bose gas in a
lattice potential is captured by the Bose-Hubbard model \cite{Jaksch_Bruder,Fisher}.
This model can be seen as an one-mode approximation that involves
the states in the first band of the optical lattice. In the tight-binding
approximation, the homogeneous Bose-Hubbard Hamiltonian for a system
of bosons in a periodic circular array with $M$ sites is given by \cite{Paraoanu}

\begin{equation}
\hat{H}_{BH}=-J\sum_{\lambda=0}^{M-1}{(\hat{a}_{\lambda}^{\dagger}\hat{a}_{\lambda+1}+\hat{a}_{\lambda+1}^{\dagger}\hat{a}_{\lambda})}+\frac{U}{2}\sum_{\lambda=0}^{M-1}{\hat{n}_{\lambda}(\hat{n}_{\lambda}-1)}.\label{eq:Bose-Hubbard}\end{equation}
In (\ref{eq:Bose-Hubbard}), $\hat{a}_{\lambda}$ and $\hat{a}_{\lambda}^{\dagger}$
are, respectively, bosonic annihilation and creation operators of
atoms on the $\lambda$th lattice site and $\hat{n}_{\lambda}\equiv\hat{a}_{\lambda}^{\dagger}\hat{a}_{\lambda}$
is the atom number operator that counts the number of atoms on the
$\lambda$th lattice site. These operators satisfy the periodic boundary
condition $\hat{a}_{\lambda}=\hat{a}_{\lambda+M}$. The first term
in Bose-Hubbard Hamiltonian is the hopping term that describes the
tunnelling of atoms among neighboring lattice sites with hopping
strength $J$ and its effect is to delocalize the atoms over the lattice.
The second term describes the inter-atomic on-site repulsion with
interaction strength $U$ whose effect is to localize
the atoms on the sites.

\subsection{The condensates}

In a condensate all atoms are in the same single-particle state and
therefore the many-body state can be written as

\begin{equation}
|c\rangle =\frac{\hat{\Gamma}^{\dagger N}}{\sqrt{N!}}\left|0\right\rangle \label{eq:condensate_state}\end{equation}
where $\hat{\Gamma}^{\dagger}$ is the creation operator of the state
occupied by the atoms

\[
\hat{\Gamma}^{\dagger}=\frac{1}{\sqrt{N}}\sum_{\lambda=0}^{M-1}{z_{\lambda}\hat{a}_{\lambda}^{\dagger}}\]
which requires that the complex parameters $z_{\lambda}$ must satisfy
the constraint

\begin{equation}
\sum_{\lambda=0}^{M-1}{|z_{\lambda}|^{2}}=N.\label{eq:constraint}
\end{equation}

The parameters $z_{\lambda}$ are determined by a variational principle
in which we minimize the mean value of $\hat{H}_{BH}$ in
the boson condensate $|c\rangle $, given by (\ref{eq:condensate_state}),
subject to the constraint (\ref{eq:constraint}). The minimization
lead us to the equation

\begin{equation}
-J(z_{\lambda+1}+z_{\lambda-1})+U\left(1-\frac{1}{N}\right)|z_{\lambda}|^{2}z_{\lambda}=\mu z_{\lambda}\label{eq:GPE_conserving}\end{equation}
which is the Gross-Pitaevskii equation for the Bose-Hubbard model
with $\mu$ being the chemical potential. A solution of this equation
that satisfy the constraint (\ref{eq:constraint}) has the general
form

\[
z_{\lambda}=\sqrt{\frac{N}{M}}e^{i\theta\lambda}\]
and, from (\ref{eq:GPE_conserving}), the chemical potential takes the explicit form 

\begin{equation}
\mu=-2J\cos\theta+\frac{UN}{M}\left(1-\frac{1}{N}\right).\label{eq:chemical_boson}\end{equation}
The possible values of $\theta$ are fixed by the periodic boundary
condition $z_{\lambda+M}=z_{\lambda}$ which requires that $\left(e^{i\theta}\right)^{M}=1$. This equation has $M$ solutions which are the $M$th roots of unity, that is, $\theta_{k}=\frac{2\pi k}{M}$ where $k$ is an integer defined on the set $\left\{ -\frac{M}{2}+1,\ldots,\frac{M}{2}\right\} $
for $M$ even or $\left\{ -\frac{\left(M-1\right)}{2},\ldots,\frac{\left(M-1\right)}{2}\right\} $
for $M$ odd. It follows from all these considerations that the bosonic atoms condense in
states with well-defined quasimomentum $\frac{2\pi k}{M}$,

\begin{equation}
\hat{A}_{k}^{\dagger}\equiv\frac{1}{\sqrt{M}}\sum_{\lambda=0}^{M-1}{e^{i\frac{2\pi k}{M}\lambda}\hat{a}_{\lambda}^{\dagger}}.\label{eq:Operador de quase-momento}
\end{equation}
These states form a single-particle basis which diagonalizes the hopping term of the Bose-Hubbard Hamiltonian. Thus, in this representation, the Hamiltonian (\ref{eq:Bose-Hubbard}) has the form

\begin{multline}
\hat{H}_{BH}=\sum_{k}{e_{k}\hat{A}_{k}^{\dagger}\hat{A}_{k}}\\+\frac{U}{2M}\sum_{\{k_{j}\}}{\delta_{M}(k_{1}+k_{2}-k_{3}-k_{4})\hat{A}_{k_{4}}^{\dagger}\hat{A}_{k_{3}}^{\dagger}\hat{A}_{k_{2}}\hat{A}_{k_{1}}}
\end{multline}
where 
\[e_{k}\equiv-2J\cos\frac{2\pi k}{M}\]
and the Kr\"{o}necker modular delta $\delta_{M}\left(k\right)$ is equal
to one if $k$ is an integer multiple of $M$ and zero otherwise.

\section{The elementary excitations\label{sec:The-Collective-Excitations}}

\subsection{The effective grand-canonical Hamiltonian and the coupling structure
of quasimomenta}

To derive the Bogoliubov-de Gennes equations we first define the shifted
operators $\hat{c}_{k}$ and $\hat{c}_{k}^{\dagger}$ by

\[
\hat{A}_{k}=z_{q}\delta_{k,q}+\hat{c}_{k},\]
where $z_{q}$ is the condensate wave function with quasimomentum $\frac{2\pi q}{M}$.
Next we write the zero temperature grand-canonical Hamiltonian $\hat{\Omega}\equiv\hat{H}_{BH}-\mu\hat{N}$,
with $\hat{N}$ denoting the number operator, as a normal order expansion with respect to the shifted operators, $\hat{c}_{k}$,

\begin{equation}
\hat{\Omega}=\hat{\omega}_{0}+\hat{\omega}_{1}+\hat{\omega}_{2}+\hat{\omega}_{3}+\hat{\omega}_{4},\label{eq:norm_order}\end{equation}
where the normal ordered operators $\hat{\omega}_{i}$ involve $i$
shifted operators. 

In the Bogoliubov theory the ground state is given by the vacuum of
the shifted operators which is a coherent state of the operator $\hat{A}_{k}$.
When we calculate the mean value of $\hat{\Omega}$ in this vacuum
state it is clear that the only contribution comes from the term $\hat{\omega}_{0}$
whose minimization with respect to $z_{q}$ lead us to equation

\begin{equation}
\left[-2J\cos\frac{2\pi q}{M}+\frac{U}{M}|z_{q}|^{2}-\mu\right]z_{q}=0.\label{eq:GPE}\end{equation}
Besides if we calculate the mean value of $\hat{N}$ in
this ground state, we get $N=\left|z_{q}\right|^{2}$.
Since this equation plus (\ref{eq:GPE}) determine $z_{q}$ except by a phase factor,
it can be taken as a real parameter. From these considerations it follows
that $z_{q}=\sqrt{N}$ and 

\begin{equation}
\mu=-2J\cos\frac{2\pi q}{M}+\frac{N}{M}U.\label{eq:chemical_coherent}\end{equation}
Notice that the expressions (\ref{eq:chemical_boson}) and (\ref{eq:chemical_coherent}),
for the chemical potential, coincide in the limit $N\gg1$. 

In the Bogoliubov theory the main hypothesis is that the majority
of atoms remains in the boson condensate state and thus the dynamics
of the system in the neighborhood of equilibrium states is described
by the quadratic term $\hat{\omega}_{2}$ since the equation (\ref{eq:GPE})
makes $\hat{\omega}_{1}$ identically zero. Thus, neglecting the third
and fourth order terms in (\ref{eq:norm_order}), we obtain the
effective grand-canonical Hamiltonian

\begin{multline}
\hat{\Omega}_{eff}^{(q)}=\hat{\omega}_{0}+\sum_{k}{(\epsilon_{k}^{(q)}+nU)\hat{c}_{k}^{\dagger}\hat{c}_{k}}\\+\frac{nU}{2}\sum_{k,k^{\prime}}{\delta_{M}(k+k^{\prime}-2q)(\hat{c}_{k}^{\dagger}\hat{c}_{k^{\prime}}^{\dagger}+\hat{c}_{k^{\prime}}\hat{c}_{k})}
\label{eq:effective}
\end{multline}
where

\begin{equation}
\epsilon_{k}^{\left(q\right)}\equiv e_{k}-e_{q}=2J\left(\cos\frac{2\pi q}{M}-\cos\frac{2\pi k}{M}\right).
\label{eq:single_particle_energy}
\end{equation}
The diagonalization of the effective Hamiltonian determines the energies and the composition of
the elementary excitations. The non-diagonal term in (\ref{eq:effective})
is the one that creates and annihilates pairs of atoms whose quasimomenta
are fixed by the condition $\delta_{M}(k+k^{\prime}-2q)=1$
which is satisfied if $(k+k^{\prime}-2q)$ is an integer
multiple of $M$, 

\begin{subequations}
\begin{equation}
k+k^{\prime}-2q=\nu\left(q\right)M.\label{eq:coupled_pairs}\end{equation}
As the quasimomenta are restricted to the first Brillouin zone, the
possible values of $\nu\left(q\right)$ are given by

\begin{equation}
\nu\left(q\right)=\begin{cases}
\begin{cases}
0 & \quad\textrm{for }M\textrm{ odd}\\
0,\:1 & \quad\textrm{for }M\textrm{ even}\end{cases} & \qquad\textrm{if }q=0\\
0,\:-\textrm{sgn}\, q & \qquad\textrm{if }q\neq0\end{cases}\end{equation}
\end{subequations}
where $\textrm{sgn}\, q$ is the signal function which is equal to
1 if $q>0$ and -1 if $q<0$. In the $\nu(q)=0$ equation both quasimomenta are in the first Brillouin zone whereas when $\nu(q)\neq0$ one of the quasimomenta is outside, the effect of $\nu(q)\neq0$ is to bring it back to the first Brillouin zone. Notice that, except in the case $q=0$ and $M$ odd, the pairs of quasimomenta coupled in (\ref{eq:effective}) are determined by two equations. 

From these equations two properties of the coupling structure of quasimomenta
emerge. The first one is the absence of ramifications, that is, one quasimomentum appears only once in the set of pairs satisfying (\ref{eq:coupled_pairs}). The second
one is that the sets of pairs of quasimomenta obtained from these
two equations are disjoint. These considerations indicate that when
$M$ is odd, we have $\frac{M-1}{2}$ pairs with distinct quasimomenta
and only one pair, $(q,q)$, with identical.
On the other hand, for $M$ even, we have $\frac{M}{2}-1$ pairs with
distinct quasimomenta and two pairs with identical, $(q,q)$
and $\left(q+\frac{M}{2}\nu(q),q+\frac{M}{2}\nu(q)\right)$,
with $\nu(q)=-1$ if $q>0$ and $\nu(q)=1$
if $q\leq0$. In the Tables \ref{tab:coupled_even} and \ref{tab:coupled_odd}, we identify the pairs of quasimomenta coupled in (\ref{eq:effective}) for $M$ even
and odd, respectively, irrespective if they differ by an exchange of the quasimomenta.

\begin{center}
\begin{table*}[ht]
\caption{The set of pairs of quasimomenta coupled in the effective grand-canonical
Hamiltonian for $M$ even.}
\label{tab:coupled_even}
\begin{centering}
{\footnotesize }\begin{tabular}{|c||cc|}
\hline 
\multicolumn{3}{|c|}{{\footnotesize $M$ even}}\\
\hline
\hline 
{\footnotesize $q<0$} & {\footnotesize $\begin{array}{c}
\nu(q)=0:\\
\nu(q)=1:\end{array}$} & {\footnotesize $\begin{array}{c}
\left(\frac{M}{2}-2|q|-1,-\frac{M}{2}+1\right),\ldots,\left(-|q|,-|q|\right),\ldots,\left(-\frac{M}{2}+1,\frac{M}{2}-2|q|-1\right)\\
\left(\frac{M}{2},\frac{M}{2}-2|q|\right),\ldots,\left(-|q|+\frac{M}{2},-|q|+\frac{M}{2}\right),\ldots,\left(\frac{M}{2}-2|q|,\frac{M}{2}\right)\end{array}$}\\
\hline 
{\footnotesize $q=0$} & {\footnotesize $\begin{array}{c}
\nu(0)=0:\\
\nu(0)=1:\end{array}$} & {\footnotesize $\begin{array}{c}
\left(\frac{M}{2}-1,-\frac{M}{2}+1\right),\ldots,\left(0,0\right),\ldots,\left(-\frac{M}{2}+1,\frac{M}{2}-1\right)\\
\left(\frac{M}{2},\frac{M}{2}\right)\end{array}$}\\
\hline 
{\footnotesize $q>0$} & {\footnotesize $\begin{array}{c}
\nu(q)=0:\\
\nu(q)=-1:\end{array}$} & {\footnotesize $\begin{array}{c}
\left(\frac{M}{2},-\frac{M}{2}+2q\right),\ldots,\left(q,q\right),\ldots,\left(-\frac{M}{2}+2q,\frac{M}{2}\right)\\
\left(-\frac{M}{2}+2q-1,-\frac{M}{2}+1\right),\ldots,\left(q-\frac{M}{2},q-\frac{M}{2}\right),\ldots,\left(-\frac{M}{2}+1,-\frac{M}{2}+2q-1\right)\end{array}$}\tabularnewline
\hline
\end{tabular}
\par\end{centering}{\footnotesize \par}
\end{table*}
\end{center}

\begin{table*}[ht]
\caption{The set of pairs of quasimomenta coupled in the effective grand-canonical
Hamiltonian for $M$ odd.}
\label{tab:coupled_odd}
\begin{centering}
{\footnotesize }\begin{tabular}{|c||cc|}
\hline 
\multicolumn{3}{|c|}{{\footnotesize $M$ odd}}\\
\hline
\hline 
{\footnotesize $q<0$} & {\footnotesize $\begin{array}{c}
\nu(q)=0:\\
\nu(q)=1:\end{array}$} & {\footnotesize $\begin{array}{c}
\left(\frac{M-1}{2}-2|q|,-\frac{M-1}{2}\right),\ldots,\left(-|q|,-|q|\right),\ldots,\left(-\frac{M-1}{2},\frac{M-1}{2}-2|q|\right)\\
\left(\frac{M-1}{2},\frac{M-1}{2}-2|q|+1\right),\ldots,\left(\frac{M-1}{2}-2|q|+1,\frac{M-1}{2}\right)\end{array}$}\\
\hline 
{\footnotesize $q=0$} & {\footnotesize $\nu(0)=0:$} & {\footnotesize $\begin{array}{c}
\left(\frac{M-1}{2},-\frac{M-1}{2}\right),\ldots,\left(0,0\right),\ldots,\left(-\frac{M-1}{2},\frac{M-1}{2}\right)\end{array}$}\\
\hline 
{\footnotesize $q>0$} & {\footnotesize $\begin{array}{c}
\nu(q)=0:\\
\nu(q)=-1:\end{array}$} & {\footnotesize $\begin{array}{c}
\left(\frac{M-1}{2},-\frac{M-1}{2}+2q\right),\ldots,\left(q,q\right),\ldots,\left(-\frac{M-1}{2}+2q,\frac{M-1}{2}\right)\\
\left(-\frac{M-1}{2}+2q-1,-\frac{M-1}{2}\right),\ldots,\left(-\frac{M-1}{2},-\frac{M-1}{2}+2q-1\right)\end{array}$}\\
\hline
\end{tabular}
\par\end{centering}{\footnotesize \par}
\end{table*}

\subsection{The Bogoliubov-de Gennes equations and elementary excitations}

The identification of the pairs of quasimomenta coupled in the effective Hamiltonian shows that it is block diagonal: $2\times2$
blocks when the quasimomenta of the pairs are identical and $4\times4$
blocks when they are distinct. In fact, (\ref{eq:effective}) can be cast into the form

\[
\hat{\Omega}_{eff}^{(q)}=\sum_{(k,k)}{\hat{h}_{2}(k,k)}+\sum_{{(k,k^{\prime})}\atop{k\neq k^{\prime}}}{{\hat{h}_{2}(k,k^{\prime})}}\]
where the first sum involves pairs with identical quasimomenta while the second distinct, with $\hat{h}_{2}(k,k)$ and $\hat{h}_{2}(k,k^{\prime})$
given explicitly by

\[
\hat{h}_{2}(k,k)=(\epsilon_{k}^{(q)}+nU)\hat{c}_{k}^{\dagger}\hat{c}_{k}+\frac{nU}{2}(\hat{c}_{k}^{\dagger}\hat{c}_{k}^{\dagger}+\hat{c}_{k}\hat{c}_{k})\]
and

\begin{eqnarray*}
\hat{h}_{2}(k,k^{\prime}) & = & (\epsilon_{k}^{(q)}+nU)\hat{c}_{k}^{\dagger}\hat{c}_{k}+(\epsilon_{k^{\prime}}^{(q)}+nU)\hat{c}_{k^{\prime}}^{\dagger}\hat{c}_{k^{\prime}}\\
& + & nU(\hat{c}_{k}^{\dagger}\hat{c}_{k^{\prime}}^{\dagger}+\hat{c}_{k^{\prime}}\hat{c}_{k}).
\end{eqnarray*}
The diagonalization of quadratic Hamiltonians of bosonic operators
is a standard problem fully explained in reference \cite{Blaizot}.
In the Appendix \ref{ap:A} we discuss briefly the procedure of this reference
to fix the notation and to present the main properties of the eigenvalue
problem. In what follows we will discuss separately the diagonalization of blocks
that involve identical and distinct quasimomenta.
\vspace{0.1cm}
\begin{itemize}
\item \textbf{Identical quasimomenta: $2\times2$ blocks}
\end{itemize}
When $M$ is odd there is only one pair with identical quasimomenta,
$(q,q)$, which is equal to the quasimomentum of the occupied
state. Solving the corresponding Bogoliubov-de Gennes equations (\ref{eq:eigen_problem})
for $\hat{h}_{2}(q,q)$, where $\mathcal{A}=\mathcal{B}=nU$, we find
that the eigenmode has zero energy and zero norm. This result is well
known and it is a consequence of the broken continuous symmetry which
arises due to violation of the atom number conservation by the Bogoliubov theory.
In this case $\hat{h}_{2}(q,q)$ can be written as

\[
\hat{h}_{2}(q,q)=\frac{\hat{\wp}^{2}}{2\chi}+\frac{nU}{2}\]
where the hermitian operator $\hat{\wp}$ is given by $\frac{1}{\sqrt{2}}(\hat{c}_{q}^{\dagger}+\hat{c}_{q})$
and the inertial parameter $\chi$ is equal to $(2nU)^{-1}$. 

On the other hand, when $M$ is even, we found two pairs with identical
quasimomenta $(q,q)$ and $\left(q+\frac{M}{2}\nu(q),q+\frac{M}{2}\nu(q)\right)$,
with $\nu(q)=-1$ if $q>0$ and $\nu(q)=1$
if $q\leq0$. The pair with quasimomentum equal to the occupied state,
$(q,q)$, gives raise to a zero-energy eigenmode as stated
before. In the case of the other pair, $\left(q+\frac{M}{2}\nu(q),q+\frac{M}{2}\nu(q)\right)$,
the Bogoliubov-de Gennes equations are given by

\begin{multline}
\left(\begin{array}{cc}
4J\cos\frac{2\pi q}{M}+nU & nU\\
nU & 4J\cos\frac{2\pi q}{M}+nU\end{array}\right)\left(\begin{array}{c}
u_{q+\frac{M}{2}\nu\left(q\right)}\\
v_{q+\frac{M}{2}\nu\left(q\right)}\end{array}\right)\\=E_{q+\frac{M}{2}\nu\left(q\right)}^{\left(q\right)}\left(\begin{array}{cc}
1 & 0\\
0 & -1\end{array}\right)\left(\begin{array}{c}
u_{q+\frac{M}{2}\nu\left(q\right)}\\
v_{q+\frac{M}{2}\nu\left(q\right)}\end{array}\right)
\end{multline}
where we have used $\epsilon_{q+\frac{M}{2}\nu(q)}^{(q)}=4J\cos\frac{2\pi q}{M}$,
from (\ref{eq:single_particle_energy}) and (\ref{eq:coupled_pairs}).
Solving these equations we find the excitation energy

\begin{equation}
E_{q+\frac{M}{2}\nu(q)}^{(q)}=\sqrt{4J\cos\frac{2\pi q}{M}\left(4J\cos\frac{2\pi q}{M}+2nU\right)}\label{eq:energy_1}\end{equation}
and the eigenvectors associated to the pair {\small $\left(E_{q+\frac{M}{2}\nu(q)}^{(q)},-E_{q+\frac{M}{2}\nu(q)}^{(q)}\right)$},
$V_{q+\frac{M}{2}\nu(q)}^{(q)}$ and $\gamma V_{q+\frac{M}{2}\nu(q)}^{(q)}$
, respectively, with {\small $V_{q+\frac{M}{2}\nu(q)}^{(q)}=\left(\begin{array}{cc}
u_{q+\frac{M}{2}\nu(q)} & v_{q+\frac{M}{2}\nu(q)}\end{array}\right)^{T}$},
where the symbol $T$ denotes the transpose of the corresponding matrix. As will be clear latter on, another useful quantity is the ratio of the components of the eigenvector given by

\begin{widetext}
\begin{equation}
\frac{v_{q+\frac{M}{2}\nu(q)}}{u_{q+\frac{M}{2}\nu(q)}}=-\frac{\left(4J\cos\frac{2\pi q}{M}+nU-\sqrt{4J\cos\frac{2\pi q}{M}\left(4J\cos\frac{2\pi q}{M}+2nU\right)}\right)}{nU}.\label{eq:rate_1}\end{equation}
\end{widetext}

\begin{itemize}
\item \textbf{Distinct quasimomenta: $4\times4$ blocks}
\end{itemize}
In this case the Bogoliubov-de Gennes equations for $\hat{h}_{2}(k,k^{\prime})$
is given by

\begin{multline}
\left(\begin{array}{cccc}
\epsilon_{k}^{(q)}+nU & 0 & 0 & nU\\
0 & \epsilon_{k^{\prime}}^{(q)}+nU & nU & 0\\
0 & nU & \epsilon_{k}^{(q)}+nU & 0\\
nU & 0 & 0 & \epsilon_{k^{\prime}}^{(q)}+nU\end{array}\right)\left(\begin{array}{c}
u_{k}\\
u_{k^{\prime}}\\
v_{k}\\
v_{k^{\prime}}\end{array}\right)
\\=E^{(q)}\left(\begin{array}{cccc}
1 & 0 & 0 & 0\\
0 & 1 & 0 & 0\\
0 & 0 & -1 & 0\\
0 & 0 & 0 & -1\end{array}\right)\left(\begin{array}{c}
u_{k}\\
u_{k^{\prime}}\\
v_{k}\\
v_{k^{\prime}}\end{array}\right).\label{eq:4x4}
\end{multline}
It is easily seen that equation (\ref{eq:4x4}) leaves uncoupled the
vectors $V_{I}^{(q)}=\left(\begin{array}{cccc}
u_{k} & 0 & 0 & v_{k^{\prime}}\end{array}\right)^{T}$ and $V_{II}^{(q)}=\left(\begin{array}{cccc}
0 & u_{k^{\prime}} & v_{k} & 0\end{array}\right)^{T}$. Therefore this equation can be reduced into two $2\times2$ eigenvalues
equations

\begin{subequations}
\begin{multline}
(I)\left(\begin{array}{cc}
\epsilon_{k}^{(q)}+nU & nU\\
nU & \epsilon_{k^{\prime}}^{(q)}+nU\end{array}\right)\left(\begin{array}{c}
u_{k}\\
v_{k^{\prime}}\end{array}\right)\\=E_{I}^{(q)}\left(\begin{array}{cc}
1 & 0\\
0 & -1\end{array}\right)\left(\begin{array}{c}
u_{k}\\
v_{k^{\prime}}\end{array}\right)\label{eq:eigen_I}\end{multline}

\begin{multline}
(II)\left(\begin{array}{cc}
\epsilon_{k^{\prime}}^{(q)}+nU & nU\\
nU & \epsilon_{k}^{(q)}+nU\end{array}\right)\left(\begin{array}{c}
u_{k^{\prime}}\\
v_{k}\end{array}\right)\\=E_{II}^{(q)}\left(\begin{array}{cc}
1 & 0\\
0 & -1\end{array}\right)\left(\begin{array}{c}
u_{k^{\prime}}\\
v_{k}\end{array}\right).\label{eq:eigen_II}\end{multline}\end{subequations}
Solving these equations we find that the excitation energies are given
by

\begin{subequations}
\begin{equation}
E_{I}^{(q)\pm}=\varepsilon_{A}^{(q)}(k,k^{\prime})\pm\sqrt{\varepsilon_{S}^{(q)}(k,k^{\prime})\left(\varepsilon_{S}^{(q)}(k,k^{\prime})+2nU\right)}\label{eq:en_1}\end{equation}
\begin{equation}
E_{II}^{(q)\pm}=-\varepsilon_{A}^{(q)}(k,k^{\prime})\pm\sqrt{\varepsilon_{S}^{(q)}(k,k^{\prime})\left(\varepsilon_{S}^{(q)}(k,k^{\prime})+2nU\right)}\label{eq:en_2}\end{equation}\end{subequations}
where

\[
\varepsilon_{A}^{(q)}(k,k^{\prime})\equiv\frac{\epsilon_{k}^{(q)}-\epsilon_{k^{\prime}}^{(q)}}{2},\qquad\varepsilon_{S}^{(q)}(k,k^{\prime})\equiv\frac{\epsilon_{k}^{(q)}+\epsilon_{k^{\prime}}^{(q)}}{2}\]
with $\epsilon_{k}^{(q)}$ given by (\ref{eq:single_particle_energy}). 

Concerning the amplitudes, we found that they satisfy the following
relations $u_{k}^{(\pm)}=v_{k}^{(\mp)}$, $u_{k^{\prime}}^{(\pm)}=v_{k^{\prime}}^{(\mp)}$,
$u_{k}^{(\pm)}=u_{k^{\prime}}^{(\pm)}$ and
$v_{k}^{(\pm)}=v_{k^{\prime}}^{(\pm)}$. These
properties allow us to identify two excitation eigenmodes whose pairs
of opposite energies are {\small $\left(E_{I}^{(q)+},-E_{I}^{(q)+}=E_{II}^{(q)-}\right)$} and {\small $\left(E_{II}^{(q)+},-E_{II}^{(q)+}=E_{I}^{(q)-}\right)$}.
The corresponding eigenvectors are $V_{I}^{(q)+}$, $\gamma V_{I}^{(q)+}=V_{II}^{(q)-}$
and $V_{II}^{(q)+}$, $\gamma V_{II}^{(q)+}=V_{I}^{(q)-}$,
whose components are {\small $V_{I}^{(q)+}=\left(\begin{array}{cccc}
u_{k}^{(+)} & 0 & 0 & v_{k^{\prime}}^{(+)}\end{array}\right)^{T}$} and {\small $V_{II}^{(q)+}=\left(\begin{array}{cccc}
0 & u_{k^{\prime}}^{(+)} & v_{k}^{(+)} & 0\end{array}\right)^{T}$}.

Thus, the diagonalization of $4\times4$ blocks gives raise
to doublets of excitation energies which are degenerate when $q=0$
and $q=\frac{M}{2}$, the last one only for $M$ even. 

One consequence of the above considerations is that the eigenmodes
belonging to a doublet have the same norm, the ratio of the amplitudes given by

\begin{widetext}\begin{equation}
\frac{v_{k^{\prime}}^{(+)}}{u_{k}^{(+)}}=\frac{v_{k}^{(+)}}{u_{k^{\prime}}^{(+)}}=-\frac{\left(\varepsilon_{S}^{(q)}(k,k^{\prime})+nU-\sqrt{\varepsilon_{S}^{(q)}(k,k^{\prime})\left(\varepsilon_{S}^{(q)}(k,k^{\prime})+2nU\right)}\right)}{nU}.\label{eq:rate_2}\end{equation}\end{widetext}

\section{Stability analysis}

\subsection{Dynamical stability}

According to the Bogoliubov theory an equilibrium state is dynamically
stable if all the excitation energies are real. The existence of at
least one complex energy is sufficient to guarantee the dynamical
instability of the corresponding condensate. In what follows we investigate
the dynamical stability of the condensates. We will analyze separately
the cases where the quasimomenta of the pair are identical and when
they are distinct.

\begin{itemize}
\item \textbf{Pair of identical quasimomenta}
\end{itemize}
As seen in the previous section, for both $M$ odd and even, there
is a zero-energy eigenmode with quasimomentum equal to the quasimomentum
of the occupied state in the condensate. This is a consequence of
the violation of atoms number conservation introduced by the Bogoliubov
approach. This mode is always present and leads to an indifferent
equilibrium which does not affect the stability of condensates.

For $M$ even there is one more pair of identical quasimomenta equal
to $k=q+\frac{M}{2}\nu(q)$, where $\nu(q)=-1$
if $q>0$ and $\nu(q)=1$ if $q\leq0$. From (\ref{eq:energy_1}) it follows that
the excitation energy is real if one of the conditions

\begin{subequations}
\begin{equation}
\cos\frac{2\pi q}{M}>0\label{eq:dy_1_s_1}
\end{equation}
or
\begin{equation}
\cos\frac{2\pi q}{M}+\frac{nU}{2J}<0\label{eq:dy_1_s_2}
\end{equation}
\end{subequations}
is obeyed. The condition (\ref{eq:dy_1_s_1}) is satisfied by the condensates
whose quasimomentum of the occupied state is in the interval
$0\leq|p|<\frac{\pi}{2}$, with $p\equiv\frac{2\pi q}{M}$, and the condition (\ref{eq:dy_1_s_2}) is satisfied when the quasimomentum is in interval $\frac{\pi}{2}<|p|\leq\pi$ such
that

\begin{equation}
r<-\cos \frac{2\pi q}{M}\label{eq:dy_single_even}\end{equation}
where $r$ is the combination of the control parameters $\frac{nU}{2J}$.

Up to now our analysis have established the domain of stability of
a particular eigenmode $k=q+\frac{M}{2}\nu(q)$. To establish
the dynamical stability of the condensate we have to analyze the behavior
of all the doublets in the control parameters space. 

\begin{itemize}
\item \textbf{Pair of distinct quasimomenta: the doublets}
\end{itemize}
From (\ref{eq:en_1}) and (\ref{eq:en_2}), the excitation energies
of the doublets corresponding to the pair $\left(k,k^{\prime}\right)$
are real if one of the two conditions

\begin{subequations}
\begin{equation}
\varepsilon_{S}^{(q)}(k,k^{\prime})>0\label{eq:dy_1}
\end{equation}
or
\begin{equation}
\varepsilon_{S}^{(q)}(k,k^{\prime})+2nU<0\label{eq:dy_1_2}
\end{equation}
\end{subequations}
is satisfied, where $\varepsilon_{S}^{(q)}(k,k^{\prime})$
is given explicitly by

\begin{equation}
\varepsilon_{S}^{(q)}(k,k^{\prime}(k))=4J\cos\frac{2\pi q}{M}\sin^{2}\frac{\pi(k-q)}{M}\label{eq:mean_single}
\end{equation}
with $k-q$ assuming the values shown in Table \ref{tab:Parametrization}.

From (\ref{eq:mean_single}) it is easily seen that the condition (\ref{eq:dy_1}) is satisfied for
condensates with quasimomenta defined in the interval $0\leq|p|<\frac{\pi}{2}$. These condensates whose quasimomentum of the occupied state is in the central region of the first Brillouin zone are always dynamically stable, independently of
the values of the control parameters. On the other hand the condition (\ref{eq:dy_1_2}) is satisfied for condensates with $\frac{\pi}{2}<|p|\leq\pi$ that obey the inequality

\begin{equation}
r<-\cos\frac{2\pi q}{M}\:\sin^{2}\frac{\pi(k-q)}{M}.\label{eq:ineq_dyn}
\end{equation}
This condition refers to stability of a particular doublet,
$(k,k^{\prime}(k))$. However we are interested
in establishing a condition that guarantee that the excitation energies
of all doublets are real. This requirement is fulfilled if we take
the minimum of the right hand side of inequality (\ref{eq:ineq_dyn})
which, according to Table \ref{tab:Parametrization}, is achieved
when $k-q=\pm1$. Thus, it follows that the condensates whose quasimomentum of the occupied state is in the last quarters of the first Brillouin zone, $\frac{\pi}{2}<|p|\leq\pi$, are dynamically stable if they obey the condition

\begin{equation}
r<-\cos \frac{2\pi q}{M}\:\sin^{2}\frac{\pi}{M}.\label{eq:cond_dy_stab}
\end{equation}
Notice that the stability of the doublets implies the stability of the equal quasimomentum pair since the inequality (\ref{eq:dy_single_even}) is contained in (\ref{eq:cond_dy_stab}).

\begin{figure}[H]
\begin{centering}
\includegraphics[scale=0.5]{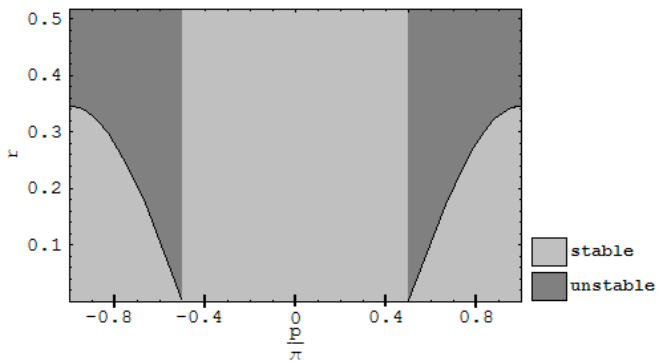}
\par\end{centering}
\caption{\label{fig:Dynamical-stability-phase}Dynamical stability phase diagram
for $M=5$ sites. Notice that $p$ assumes only discrete values $0$,
$\pm0.4\pi$, $\pm0.8\pi$ indicated in the figure. The
condensates with quasimomenta $0$ and $\pm0.4\pi$, which are in
the central region of the first Brillouin zone, are always dynamically
stable. On the other hand the dynamical stability of the condensates
with quasimomenta $\pm0.8\pi$, which are in the last quarters of
the first Brillouin zone, depends on the control parameters. The continuous
curve is only to guide the eye. }
\end{figure}

In the Table \ref{tab:Dynamical-stability}
we summarize our findings and a dynamical stability phase diagram
is shown in Figure \ref{fig:Dynamical-stability-phase} for $M=5$
sites. We disregard the condensates with quasimomentum $|p|=\frac{\pi}{2}$,
the reason being that we cannot find a Bogoliubov transformation that
diagonalizes the effective grand-canonical Hamiltonian since all eigenvectors
of the Bogoliubov-de Gennes equation have zero norm.

\begin{table}[H]
\caption{\label{tab:Dynamical-stability}Dynamical stability of the condensates.}
\begin{centering}
\begin{tabular}{|c|c|}
\hline 
\multicolumn{2}{|c|}{Dynamical Stability}\\
\hline
\hline 
$0\leq|p|<\frac{\pi}{2}$ & always stable\\
\hline 
$\frac{\pi}{2}<|p|\leq\pi$ & stable if $r<-\cos p\:\sin^{2}\frac{\pi}{M}$\\
\hline
\end{tabular}
\par\end{centering}
\end{table}

When $r=0$ all the condensates are dynamically stable. Actually they
are eigenstates of the Bose-Hubbard Hamiltonian when $U=0$. When
$r$ starts to increase the condensates with the smallest $|p|$ in the set $\frac{\pi}{2}<|p|\leq\pi$ start to become unstable until $r>\sin^{2}\frac{\pi}{M}$, when all the
condensates in the last quarters of the first Brillouin zone are unstable.

Recent experiments \cite{Campbell} have shown that the onset of dynamical instability occur when two atoms in the condensate can elastically scatter into a final state where they have quasimomenta different from $p=\frac{2\pi q}{M}$. To point out the relevance of our analysis to this matter, we cast the stability condition (\ref{eq:dy_1}) into the form \cite{Wu}
\begin{equation}
2e_{q}<e_{k}+e_{k^{\prime}}.
\end{equation}
This condition reveals that in the condensates whose quasimomentum of the occupied state is in the central region of the first Brillouin zone, pairs of atoms cannot elastically scatter into two different quasimomentum states $\frac{2\pi k}{M}$ and $\frac{2\pi k^{\prime}}{M}$.

For the condensates whose quasimomentum of the occupied state is in the last quarters of the first Brillouin zone the stability condition (\ref{eq:dy_1_2}) takes the form
\begin{equation}
2e_{q}>e_{k}+e_{k^{\prime}}+2nU.
\end{equation}
Notice that in this case the quantity $2e_{q}-e_{k}-e_{k^{\prime}}$ is positive and has a lower limit equal to $2nU$ which implies that two atoms can inelastically scatter with the excess of energy being transferred to the other atoms in the system. However, notice that in the thermodynamical limit all these condensates are dynamically unstable.

\subsection{Energetic stability}

In the framework of the Bogoliubov theory an equilibrium state is
energetically unstable if there is at least one negative excitation
energy. Recall that to an elementary excitation we associate a pair
of eigenvectors with opposite eigenvalues $(E,-E)$ and opposite sign of the norm, where the
value of the excitation energy is the eigenvalue of the eigenvector with a positive norm. 

We can distinguish two mechanisms of energetic instability. One is ``crossing" that occurs when a positive excitation energy
vanishes and changes the sign and the second is ``no-crossing" where we
always have a negative excitation energy. The difference between
these two mechanisms is that ``crossing" depends on the control
parameters whereas ``no-crossing" does not.

We restrict our analysis of energetic stability to dynamically stable
condensates whose excitation energies are all real, a necessary condition
of energetic stability. Our procedure to establish the energetic stability
of the condensates is, for each eigenmode, to identify the eigenvector
with positive norm and the sign of the corresponding eigenvalue. The sign
of the norm depends on the size of the ratio $\frac{v}{u}$. If $0<\left|\frac{v}{u}\right|<1$
the norm is positive and if $\left|\frac{v}{u}\right|>1$ the norm
is negative. Following what we have done in the case of dynamical stability, we will discuss
separately the case of identical and distinct pairs of quasimomenta.

\begin{itemize}
\item \textbf{Pair of identical quasimomenta}
\end{itemize}
As pointed out before the zero-energy eigenmode does not affect the
stability of the condensates. Therefore we are left
with the pair $k=q+\frac{M}{2}\nu(q)$, with $\nu(q)=-1$
if $q>0$ and $\nu(q)=1$ if $q\leq0$, that exists only for $M$ even. By inspection of
the equation (\ref{eq:rate_1}) we find that

\begin{flushleft}
a) if $\cos\frac{2\pi q}{M}>0$, then $V_{q+\frac{M}{2}\nu(q)}^{(q)}$
has a positive norm;
\par\end{flushleft}

\begin{flushleft}
b) if $\cos\frac{2\pi q}{M}+\frac{nU}{2J}<0$, then $V_{q+\frac{M}{2}\nu(q)}^{(q)}$
has negative norm.
\par\end{flushleft}

Since in both cases the corresponding energy $E_{q+\frac{M}{2}\nu(q)}^{(q)}$
is always positive we conclude that in case a), where $0\leq|p|<\frac{\pi}{2}$,
this particular eigenmode is always energetically stable. On the other
hand, in case b), where $\frac{\pi}{2}<|p|\leq\pi$, it
is always energetically unstable.

\begin{itemize}
\item \textbf{Pair of distinct quasimomenta: the doublets}
\end{itemize}
Inspection of equation (\ref{eq:rate_2}) shows that,

\begin{flushleft}
a) if $\varepsilon_{S}^{(q)}(k,k^{\prime})>0$,
then $V_{I}^{(q)+}$ and $V_{II}^{(q)+}$ have
positive norm;
\par\end{flushleft}

\begin{flushleft}
b) if $\varepsilon_{S}^{(q)}(k,k^{\prime})+2nU<0$,
then $V_{I}^{(q)+}$ and $V_{II}^{(q)+}$ have
negative norm.
\par\end{flushleft}

Notice from (\ref{eq:en_1}) and (\ref{eq:en_2}) that the sign of the energies $E_{I}^{(q)+}$ and
$E_{II}^{(q)+}$ depends on the term $\varepsilon_{A}^{(q)}(k,k^{\prime})$, given explicitly by

\begin{equation}
\varepsilon_{A}^{(q)}(k,k^{\prime}(k))=2J\sin\frac{2\pi q}{M}\sin\frac{2\pi(k-q)}{M}\label{eq:difference}
\end{equation}
for all the pairs of quasimomenta $(k,k^{\prime}(k))$. According to Table \ref{tab:Parametrization}, we see that this term is semi-negative definite, 

\[
\varepsilon_{A}^{(q)}(k,k^{\prime}(k))\leq0.\]
This property implies that $E_{II}^{(q)+}$
is always positive, whereas $E_{I}^{(q)+}$ can change its
sign.

In the case a), where $p$ is defined in the range $0\leq|p|<\frac{\pi}{2}$,
the condition for the $(k,k^{\prime}(k))$
mode to be energetically stable is $E_{I}^{(q)+}>0$ which
can be cast into the form

\[
\varepsilon_{S}^{(q)}(k,k^{\prime})\left(\varepsilon_{S}^{(q)}(k,k^{\prime})+2nU\right)-\left[\varepsilon_{A}^{(q)}(k,k^{\prime})\right]^{2}>0.\]
Inserting the expressions (\ref{eq:mean_single}) and (\ref{eq:difference})
in the above inequality we obtain

\[
\cos^{2}\frac{2\pi q}{M}+r\cos\frac{2\pi q}{M}>\cos^{2}\frac{\pi(k-q)}{M}.\]
However we are interested in a condition that guarantee the stability
of all the doublets. This requirement is satisfied if we take the
maximum of the right hand side, which is achieved at $k-q=\pm1$. Thus,
the condensates such that $0\leq|p|<\frac{\pi}{2}$ are energetically
stable if they obey the condition

\[
\cos^{2}\frac{2\pi q}{M}+r\cos \frac{2\pi q}{M}>\cos^{2}\frac{\pi}{M},\]
which can be written in a more convenient way as

\begin{equation}
\sin\left(\frac{2\pi q}{M}-\frac{\pi}{M}\right)\sin\left(\frac{2\pi q}{M}+\frac{\pi}{M}\right)<r\cos \frac{2\pi q}{M}.\label{eq:energetic}\end{equation}
Thus, in this case, the route to energetic instability is the ``crossing" mechanism.

In the case b) where the condensates are such that
$\frac{\pi}{2}<|p|\leq\pi$, one of the energies of the
doublets are always negative. To see this, note that $E_{II}^{(q)+}$
is always positive and the corresponding eigenvector, $V_{II}^{(q)+}$
, has negative norm. Thus, by the Bogoliubov criteria, the excitation
energy of this mode is $-E_{II}^{(q)+}=E_{I}^{(q)-}$
which is strictly negative, independently of the control parameters. Therefore, we conclude that the
condensates whose quasimomentum of the occupied state is in the last quarters of the first Brillouin zone are always energetically unstable, in this case through the ``no-crossing" mechanism. In the Table \ref{tab:Energetic-stability} we
summarize our findings.

\begin{table}[H]
\caption{\label{tab:Energetic-stability}Energetic stability of condensates.}
\begin{centering}
\begin{tabular}{|c|c|}
\hline 
\multicolumn{2}{|c|}{Energetic Stability}\\
\hline
\hline 
$0\leq|p|<\frac{\pi}{2}$ & stable if $\sin\left(p-\frac{\pi}{M}\right)\sin\left(p+\frac{\pi}{M}\right)<r\cos p$\\
\hline 
$\frac{\pi}{2}<|p|\leq\pi$ & always unstable\\
\hline
\end{tabular}
\par\end{centering}
\end{table}

\begin{figure}[H]
\begin{centering}
\includegraphics[scale=0.5]{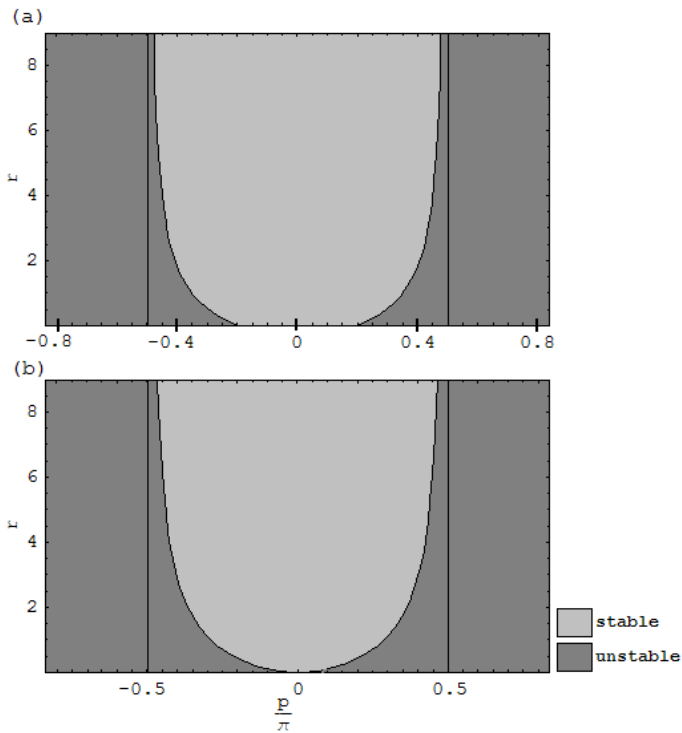}
\par\end{centering}

\caption{\label{fig:Energetic-stability-phase}Energetic stability phase diagram (a) for $M=5$ sites and (b) in
the thermodynamical limit $M\gg1$, $n=\frac{N}{M}$ fixed. In (a), since $r$ is a semi-positive definite parameter, we did
not plot the curve that defines the boundary of energetic stability in the interval $0\leq |p|\leq\frac{\pi}{M}$ where
$r$ is negative. Notice that, in (a), $p$ assumes only discrete values $0$,
$\pm0.4\pi$, $\pm0.8\pi$ indicated in the figure and the continuous curve is only to guide the eye. The two vertical
continuous lines define the region where the condensates are always
dynamically stable, independently of control parameters.}

\end{figure}

In the Figure \ref{fig:Energetic-stability-phase} we present the energetic stability phase diagram for $M=5$ sites and in the thermodynamical limit, $M\gg1$, $n=\frac{N}{M}$
fixed. We plot the curve that defines the boundary of energetic stability only for $r\geq0$. This curve is confined in the central region of the first Brillouin zone
where the condensates are always dynamically stable, independently
of the control parameters. In the last quarters, the condensates are
always energetically unstable.

One question that we can address is to determine the interval of control
parameters in which there are metastable current carrying condensates, that is, dynamically and energetically stable condensates where the occupied states have a finite quasimomentum. These superflow states correspond to local minima of the energy and they are candidates to present superfluid motion. To find this interval notice that we can define a critical
value $r$, say $r_{\textrm{st}}$, such that
for $r\geq r_{\textrm{st}}$ the condensates with $0<|p|<\frac{\pi}{2}$
are all metastable. If $p_{\textrm{max}}$ is the highest value of
$p$ defined in the interval $0<|p|<\frac{\pi}{2}$, from
(\ref{eq:energetic}) the critical value $r_{\textrm{st}}$ is given
explicitly by

\begin{equation}
r_{\textrm{st}}=\frac{\sin\left(p_{\textrm{max}}-\frac{\pi}{M}\right)\sin\left(p_{\textrm{max}}+\frac{\pi}{M}\right)}{\cos p_{\textrm{max}}}.\label{eq:rst}\end{equation}
We can also define another critical value $r$, say $r_{\textrm{unst}}$, such
that for $r<r_{\textrm{unst}}$ there is no metastability. When this occur only the $p=0$ condensate is stable. From this consideration follows that

\[
r_{\textrm{unst}}=\frac{\sin\left(p_{\textrm{min}}-\frac{\pi}{M}\right)\sin\left(p_{\textrm{min}}+\frac{\pi}{M}\right)}{\cos p_{\textrm{min}}}.\]
Since $\left|p_{\textrm{min}}\right|=\frac{2\pi}{M}$, $r_{\textrm{unst}}$
is given by

\begin{equation}
r_{\textrm{unst}}=\frac{\sin\left(\frac{3\pi}{M}\right)\sin\left(\frac{\pi}{M}\right)}{\cos\frac{2\pi}{M}}.\label{eq:runst}\end{equation}
When $r$ increases continuously from zero there is a hierarchy in
the appearance of these superflow states, starting at $r=r_{\textrm{unst}}$
where pairs of degenerate condensates with quasimomentum $p$ and $-p$, beginning with $p=\frac{2\pi}{M}$, become metastable up to $r\geq r_{\textrm{st}}$ when all states are metastable.

\section{Summary and conclusions}

In this paper we use the Bose-Hubbard model and the Bogoliubov theory to investigate the properties
of ultra cold bosonic atoms loaded in a periodic ring with $M$ sites. First we derive and solve the
Gross-Pitaevskii equation of the model and from the analysis of the
solutions we show that the atoms condense in states with well-defined
quasimomentum whose values are the $M$th roots of unit, restricted to the first Brillouin zone. Thus, besides the usual zero quasimomentum condensate, we have equilibrium states with non-zero quasimomentum which correspond to current carrying condensates \cite{Bloch}. These states with well-defined quasimomentum form a basis that diagonalize the hopping term of the Bose-Hubbard Hamiltonian.

Following the Bogoliubov theory we derive the effective grand-canonical Hamiltonian, quadratic in the shifted operators, whose diagonalization gives the energies and the composition of the elementary excitations. A detailed analysis of the coupling structure in the effective grand-canonical Hamiltonian shows
that only pairs of identical and distinct quasimomenta are coupled.
An immediate consequence of this coupling structure
is that the effective Hamiltonian is block diagonal: $2\times2$ blocks
when the quasimomenta of the pairs are identical and $4\times4$ blocks
when they are distinct. The diagonalization of the $2\times2$ blocks shows that the pair $(q,q)$ with quasimomentum equal to the quasimomentum of the occupied state in the condensate gives raise to the zero-energy eigenmode. On the other, the diagonalization of the $4\times4$ blocks gives raise to doublets of excitation energies which are degenerate when $q=0$ and, for $M$ even, $q=\frac{M}{2}$. This shows that when $q\neq 0$ the excitation spectrum has a two-branch structure. We have also found, by inspection of the excitation energies (\ref{eq:en_1}) and (\ref{eq:en_2}), that the phonon limit is achieved when the relative quasimomentum $l-p$ goes to zero, with $l=\frac{2\pi k}{M}$ being the quasimomentum of the excitation. Indeed, in this limit it follows that $E_{I,II}=c_{I,II}|l-p|$ with two different sound velocities: $c_{I}^{+}=2J(-\sin p+\sqrt{r\cos p})$ and $c_{II}^{+}=2J(\sin p+\sqrt{r\cos p})$. These properties are signatures of the finite size of the quasimomentum of the occupied state in the condensates.

Our stability analysis shows that the condensates in the central region
of the first Brillouin zone, $0\leq|p|<\frac{\pi}{2}$,
are always dynamically stable whereas the dynamical stability of the condensates in the last quarters, $\frac{\pi}{2}<|p|\leq\pi$, depends on the control parameters. When $r$ increases from zero, these condensates start to become unstable beginning with the one of smallest $|p|$ ending up with the instability of all condensates, when $r>\sin^{2}\frac{\pi}{M}$.

Concerning the energetic stability, we found that the condensates in the last quarters of the first Brillouin zone are always unstable whereas the energetic stability of the condensates in the central region depends on the control parameters. We show that when $r<r_{\textrm{unst}}$ only the $p=0$ condensate is stable. However, when $r>r_{\textrm{unst}}$, pairs of degenerate condensates with non-zero quasimomentum, $p$ and $-p$, start to become energetically stable ending up with all the condensates in the central region of the first Brillouin zone being energetically stable, when $r>r_{\textrm{st}}$.

As discussed in the paper, metastable current carrying condensates are candidates to present superfluid motion. A metastable state is both dynamically and energetically stable, consequently a local minimum of the energy. Our analysis shows that the Bogoliubov theory predicts that there is an interval in the control parameters space where metastable current carrying condensates exist. The number of these states increases with $r$ and they are restricted to the central region of the first Brillouin zone.

\begin{acknowledgments}
ETDM and EJVP would like to acknowledge financial support from FAPESP and CNPq. 
\end{acknowledgments}

\appendix

\section{\label{ap:A}The diagonalization of a quadratic Hamiltonian of bosonic operators}

A general quadratic Hamiltonian of bosonic operators is given by

\[
\hat{H}=\sum_{i,j}{\mathcal{A}_{ij}\hat{c}_{i}^{\dagger}\hat{c}_{j}}+\frac{1}{2}\sum_{i,j}{(\mathcal{B}_{ij}\hat{c}_{i}^{\dagger}\hat{c}_{j}^{\dagger}+\mathcal{B}_{ij}^{\star}\hat{c}_{j}\hat{c}_{i})}\]
where $\hat{c}_{i}^{\dagger}$ and $\hat{c}_{i}$ are creation and
annihilation bosonic operators, $\mathcal{A}_{ij}$ and $\mathcal{B}_{ij}$ are elements
of an hermitian and symmetric matrices, respectively. The diagonalization
of this quadratic Hamiltonian consists in finding a canonical transformation
to quasiparticle operators,

\[
\hat{b}_{n}^{\dagger}\equiv\sum_{r}{(u_{r}^{n}\hat{c}_{r}^{\dagger}-v_{r}^{n}\hat{c}_{r})},\]
such that $\hat{H}$, when written in terms of this operators, takes
the form of a system of non-interacting quasiparticles. From this
constraint it follows that

\begin{equation}
\left(\begin{array}{cc}
\mathcal{A} & \mathcal{B}\\
\mathcal{B}^{\star} & \mathcal{A}^{\star}\end{array}\right)\left(\begin{array}{c}
u^{n}\\
v^{n}\end{array}\right)=E_{n}\left(\begin{array}{cc}
1 & 0\\
0 & -1\end{array}\right)\left(\begin{array}{c}
u^{n}\\
v^{n}\end{array}\right)\label{eq:eigen_problem}\end{equation}
which is the matrix form of Bogoliubov-de Gennes equations that determines
the excitation energies $E_{n}$ and the composition of the elementary
excitations $V^{n}=\left(\begin{array}{cc}
u^{n} & v^{n}\end{array}\right)^{T}$, with $T$ denoting the transpose of the corresponding matrix.
When the eigenvalues $E_{n}$ are real, they appear in pairs $\left(E_{n},-E_{n}\right)$ with opposite sign of the norm, $\sum_{r}{(|u_{r}^{n}|^{2}-|v_{r}^{n}|^{2})}$. In fact, if $V^{n}$ is an eigenvector with the eigenvalue
$E_{n}$ then $\gamma V^{n\star}$, with

\[
\gamma\equiv\left(\begin{array}{cc}
0 & 1\\
1 & 0\end{array}\right),\]
is an eigenvector with the opposite eigenvalue, $-E_{n}$, and opposite sign of the norm. The excitation energy is identified with the eigenvalue whose eigenvector has a positive norm.

\section{\label{ap:B}Parametrization of the pairs of distinct quasimomenta}

As seen before the pairs of distinct quasimomenta, $\left(k,k^{\prime}\right)$,
coupled in the effective grand-canonical Hamiltonian identify the
doublets that compose the excitation spectrum of the condensates. However $k$ and $k^{\prime}$ are not independent since they are related by equation (\ref{eq:coupled_pairs}). Therefore we need one parameter to identify the doublets. Our choice was the relative quasimomentum, $\frac{2\pi}{M}(k-q)$, which leads to an ordered and one-to-one parametrization of the doublets. This can be easily seen noticing that we can cast the equation (\ref{eq:coupled_pairs}) into the form

\[(k-q)+(k^{\prime}-q)=\nu\left(q\right)M,\]
and, from it, follows the parametrization shown in the Table \ref{tab:Parametrization}.

\begin{table}[H]
\begin{centering}
\caption{\label{tab:Parametrization}Identification of the doublets in the
excitation spectrum by the relative quasimomenta $\frac{2\pi}{M}(k-q)$ of the atoms in the condensate.}
{\footnotesize }\begin{tabular}{|c||c|c|}
\hline 
{\footnotesize Condensate} & \multicolumn{2}{c|}{\footnotesize Doublets ($k-q$)}\\
\cline{2-3}
& {\footnotesize $M$ odd} & {\footnotesize $M$ even}\\
\hline\hline
{\footnotesize $q\leq0$} & {\footnotesize $1\leq k-q\leq \frac{M-1}{2}$}& {\footnotesize $1\leq k-q\leq\frac{M}{2}-1$}\\
\hline 
{\footnotesize $q>0$} & {\footnotesize $-\frac{M-1}{2}\leq k-q\leq-1$}& {\footnotesize $-\frac{M}{2}+1\leq k-q\leq-1$}\\
\hline
\end{tabular}
\par\end{centering}{\footnotesize \par}
\end{table}

This parametrization can be extended to include the pairs of identical quasimomenta. Indeed, $k-q=0$ for the $(q,q)$ pair and $k-q=\nu(q)\frac{M}{2}$ for the $\left(q+\nu(q)\frac{M}{2},q+\nu(q)\frac{M}{2}\right)$ pair.

\end{document}